\documentclass{ws-procs9x6}
\newcommand{\pt} {p$_{_{T}}$}

\begin{document}

%
%


\title{Identified particle measurements at large transverse momenta
from p+p to Au+Au collisions at RHIC.}

\author{R.~S.~HOLLIS\footnote{\uppercase{F}or the \uppercase{STAR C}ollaboration.}}

\address{Department of Physics\\
University of Illinois at Chicago\\
845 W. Taylor Street (MC 273), \\ 
Chicago, IL 60607, USA\\ 
E-mail: rholli3@uic.edu}

\maketitle

\abstracts{
Measurements of various particle species over an extended momentum
range provide a sensitive experimental tool for investigating particle
production mechanisms in hadronic collisions.  Comparison of the spectral
shapes from different collision centralities measured with the STAR
detector at RHIC allows one to study the interplay of soft and hard
particle production for mesons and investigate various baryon-meson
effects.  Systematic studies of identified particle spectra for various
colliding systems and different incident energies provide additional
insights toward the interplay between fragmentation and non-fragmentation
contributions to the particle production.  In these proceedings we present
a systematic study of transverse momentum spectra for charged pions,
protons and antiprotons from Au+Au and Cu+Cu data at $\sqrt{s_{NN}}=200$
and 62.4~GeV as a function of collision centrality.  We compare those
measurements with p+p and d+Au data, investigating the system effects
on energy loss.}

\section{Introduction}

Spectral measurements yield valuable information on the state of matter
produced in collisions of heavy nuclei in the relativistic energy regime.
In these proceedings the importance of identifying particle species from
low- (\pt$\sim0.4$~GeV$/c$) to high-\pt~(up to 10~GeV$/c$) are
discussed in the context of partonic propagation through the hot, dense
medium produced in these collisions at the Relativistic Heavy-Ion Collider
(RHIC).

Hard partonic interactions, occurring in the early stages of the collision,
are known to produce high momentum particles resulting from quark or gluon
scattering and subsequent fragmentation\footnote{from p+p collisions}.
These hard scatterings still occur in heavy-ion collisions, although the
partons (through their final products) are found to undergo modification
upon propagating the hot, dense medium created during the collision.

High-\pt~particles are a valuable probe that could help to understand
parton fragmentation and their interactions with the created hot, dense
medium.  Understanding of modifications to high-\pt~particle distributions
can lead to qualitative conclusions on the energy loss mechanisms in the
medium.  As the unmodified (vacuum fragmentation) distribution of
high-\pt~particles is known from elementary p+p collision
data and is well described by pQCD calculations, comparative analysis
provides a distinct advantage for high-\pt~particles.

Various particle species at intermediate- to high-\pt~are expected to
have very different contributions from quark and gluon jet
fragmentation\cite{cite:STAR_Id200dAupp,cite:STAR_Id200AuAu}.
Specifically, intermediate-\pt~protons come predominantly from gluon jets,
due to their softer fragmentation function.  In fragmentation, the
majority of pions originate from quark jets by string breaking into
quark-antiquark pairs in the intermediate-\pt~range.  One can then study
color-charge differences of energy loss for quark and gluon jets via the
identified particle spectra.

It is expected that effects of jet-medium interactions on the final
particle distributions are strongly dependent on the size of that medium.
Spectral shapes\cite{cite:STAR_SpectralSuppression} and azimuthal
correlations\cite{cite:STAR_BackToBack} in the most central
(or fully overlapping) collisions are found to be subjected to the highest
modification, whilst d+Au or peripheral Au+Au collisions are relatively
unmodified.  The addition of the smaller Cu+Cu collision system augments
the information in this regard, bridging the gap between d+Au and
peripheral Au+Au collisions.

In these proceedings, the centrality, system size and \pt~dependence
of the energy loss is explored in the context of the nuclear modification
factor and the magnitude of relative baryon enhancement found in such
collisions.

\section{Detector and Methods}

The identified particle spectra presented here utilize data collected by
the STAR detector at RHIC over the past 6 years. The main apparatus is
the time projection chamber, TPC, which covers the full azimuth over the
rapidity range $|\eta|<1.8$.  The ionization energy loss in the TPC is
used to identify $\pi^{\pm}$, K$^{\pm}$ and protons and antiprotons in
the range $0.3$~GeV$/c < $\pt$ < 1.2$~GeV$/c$, a kinematic region where
particles of different masses are clearly separated in $dE/dx$.  For the
high-\pt~regime (\pt$>2.5$~GeV$/c$), the \pt~dependence of the relativistic
rise of the ionization energy loss is used to statistically separate the
particles.  Data from these two \pt~regions are augmented by additional
coverage afforded by the prototype Time of Flight detector, which
identifies particles in the range $0.2$~GeV$/c < $\pt$ < 3.0$~GeV$/c$.
More details of the analysis techniques used can be found
in Refs.~\cite{cite:STAR_Id200dAupp,cite:STAR_ToF}.

\section{Results and Discussion}

\subsection{Nuclear Modification Factor}
\begin{figure}[ht]
\begin{minipage}[t]{.45\textwidth}
\centerline{\epsfxsize=2.4in\epsfbox{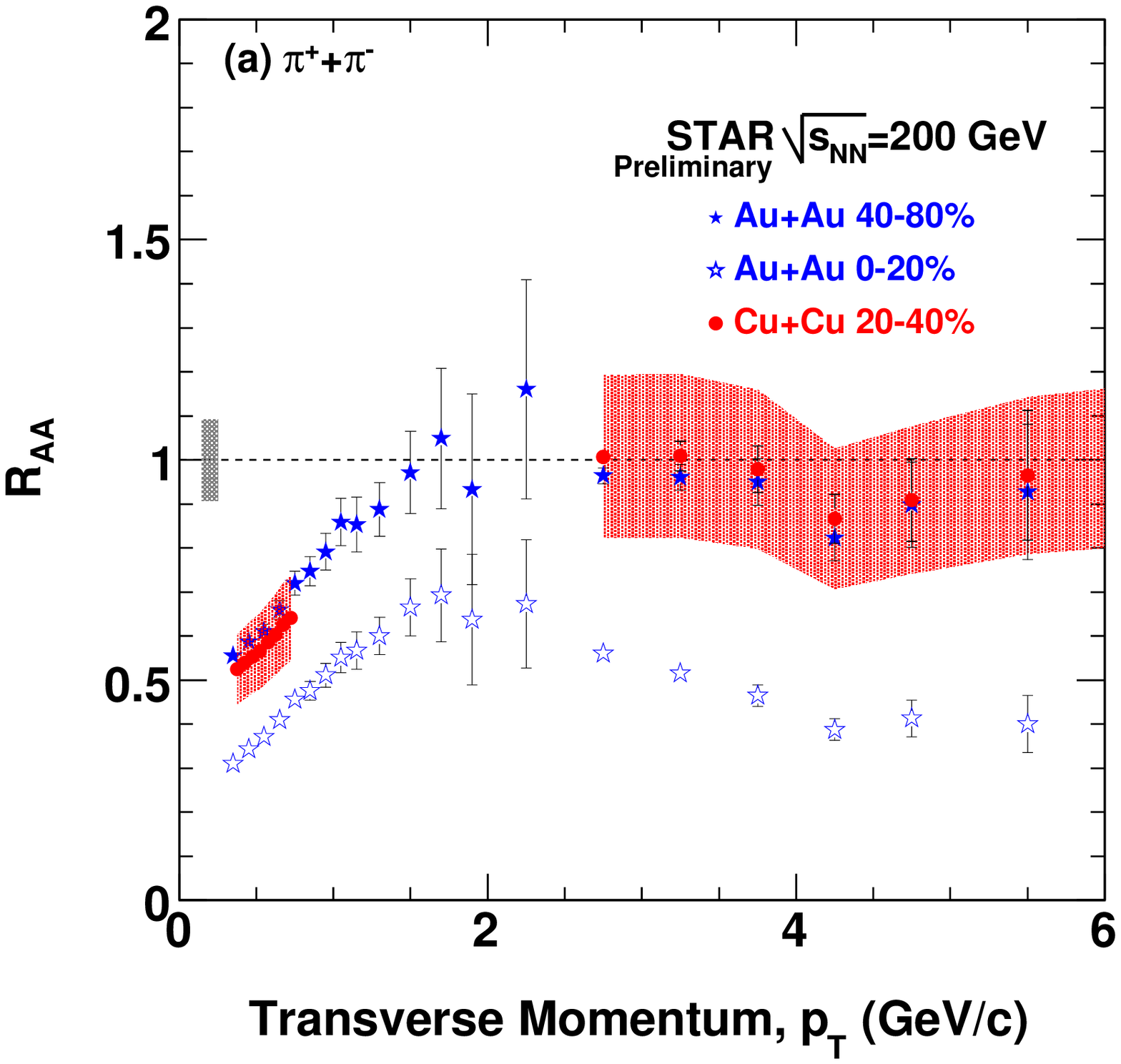}}
\end{minipage}
\hspace{20pt}
\begin{minipage}[t]{.45\textwidth}
\centerline{\epsfxsize=2.4in\epsfbox{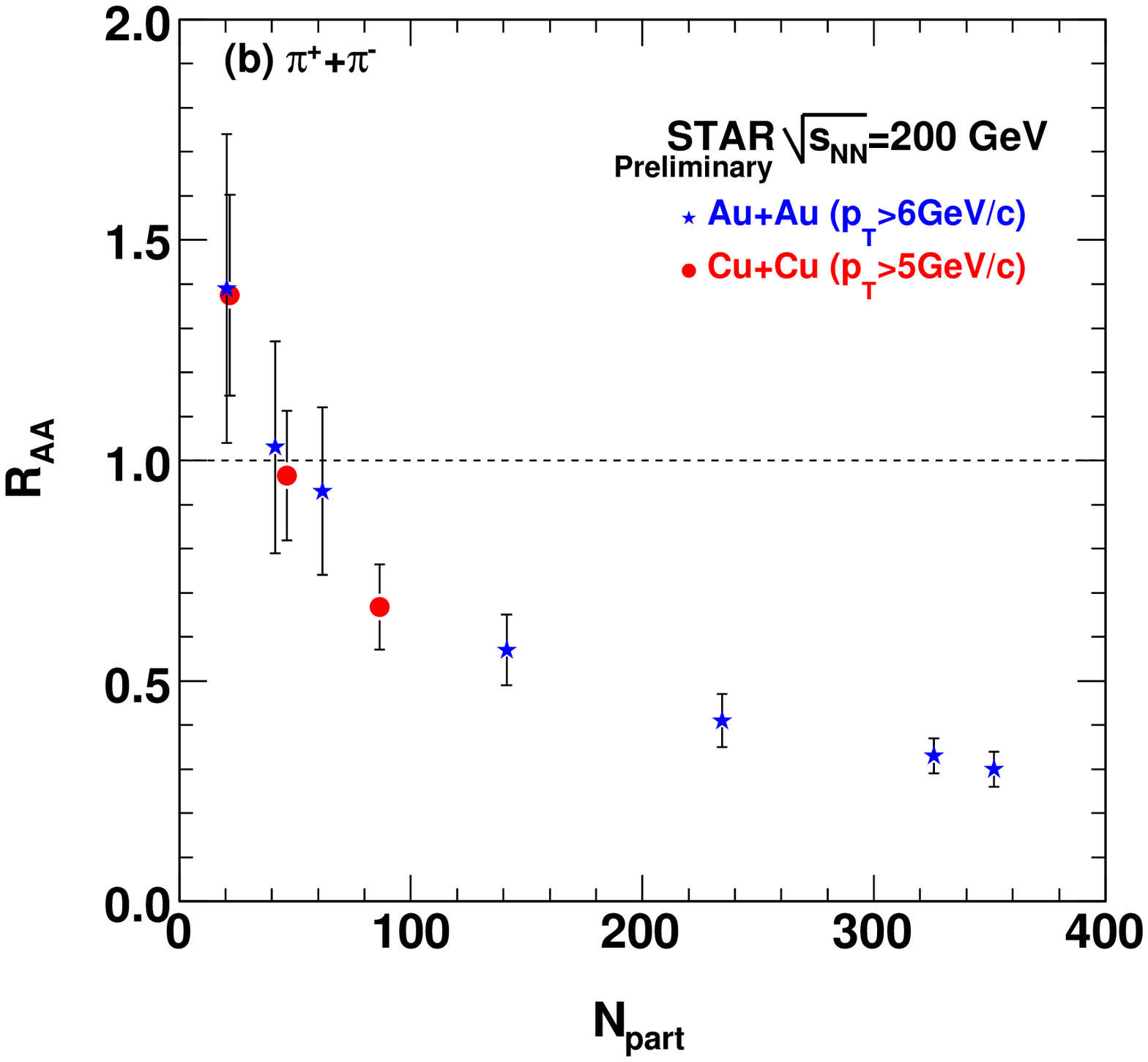}}
\end{minipage}
\caption{Panel (a) shows the transverse momentum dependence of the nuclear
modification factor for $\pi^{+}+\pi^{-}$ in Au+Au collisions at
$\sqrt{s_{NN}}=200$~GeV for a central and a peripheral centrality bin. 
For comparison, Cu+Cu data at the same energy is shown for an equivalent
$N_{part}$ bin to the peripheral Au+Au.  The shaded band over the data
represents the systematic uncertainty in the Cu+Cu data, the grey shaded
band illustrates the scale uncertainty from $N_{coll}$.  Panel (b) shows
the centrality dependence of the energy loss for the same data, averaged
for \pt$>6$~GeV$/c$.
\label{fig:PionRAA}}
\end{figure}

Modification to the spectral distributions due to the created medium
in the heavy-ion collisions are most directly seen in the ratio with
an appropriate reference spectra from p+p collisions.  Such a ratio,
scaled by the number of binary collisions ($N_{bin}$), is commonly
referred to as the Nuclear Modification Factor (denoted as R$_{AA}$).
As can be seen from Fig.~\ref{fig:PionRAA}a, $\pi^{+}+\pi^{-}$ production is
significantly suppressed at high-\pt~for the most central data compared
to the expectation from the binary scaled p+p reference.  For more
peripheral events, the modification is smaller and is found to
extrapolate back smoothly to the p+p reference (Fig.~\ref{fig:PionRAA}b).
These features of the data have been explained by induced energy
loss of the partons traversing the hot, dense medium.  Such an effect
reproduces the centrality dependence of the partonic energy loss found
in data\cite{cite:Theory_Vitev}.  Cu+Cu data add to this systematic study
of system size effects of energy loss in heavy ion collisions.  A
comparison of the same average system size (or number of participants)
reveals little difference between the two systems, see Fig.~\ref{fig:PionRAA}.

\subsection{Baryon to meson enhancement}
\begin{figure}[ht]
\begin{minipage}[t]{.45\textwidth}
\centerline{\epsfxsize=2.4in\epsfbox{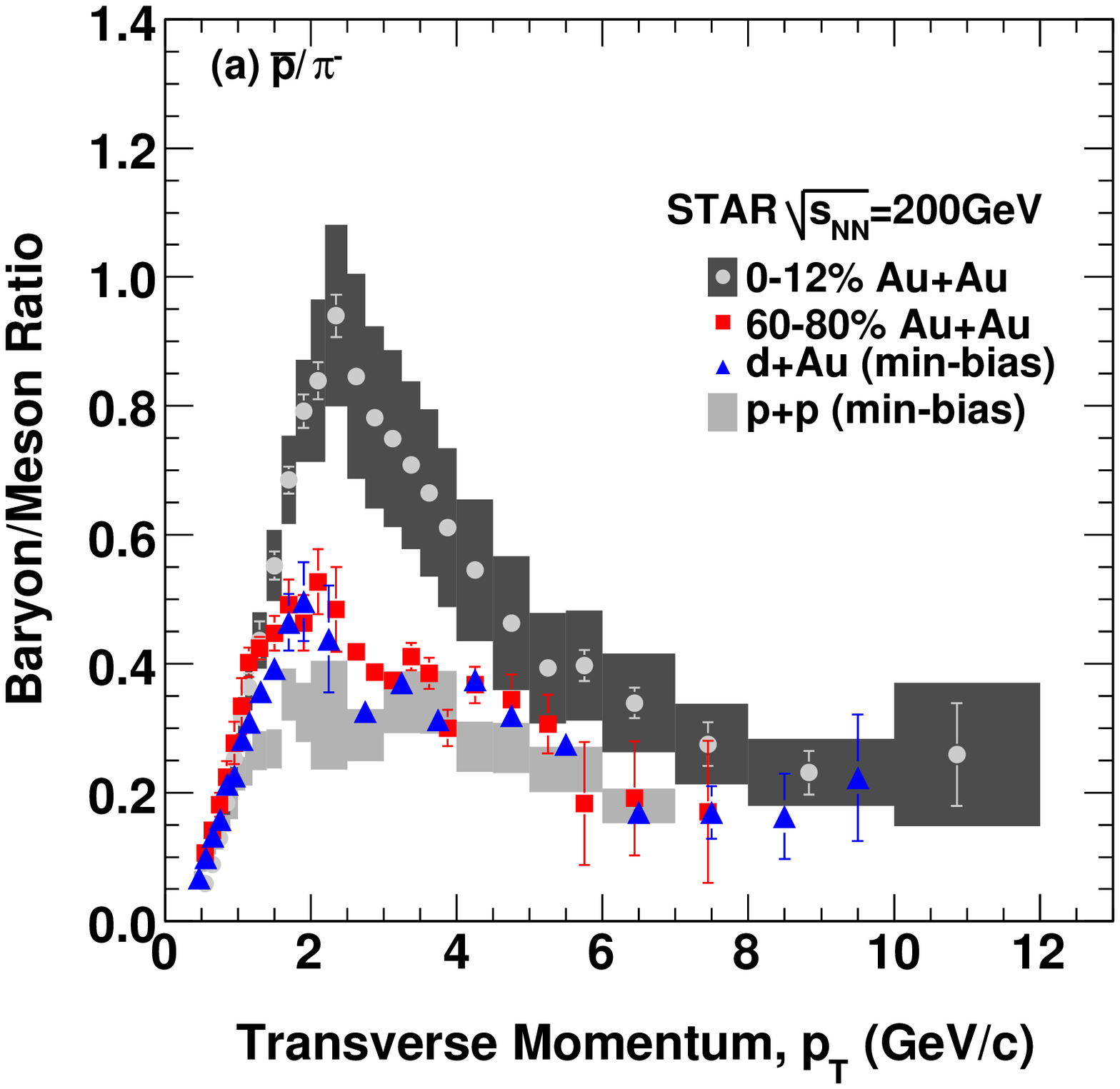}}
\end{minipage}
\hspace{20pt}
\begin{minipage}[t]{.45\textwidth}
\centerline{\epsfxsize=2.4in\epsfbox{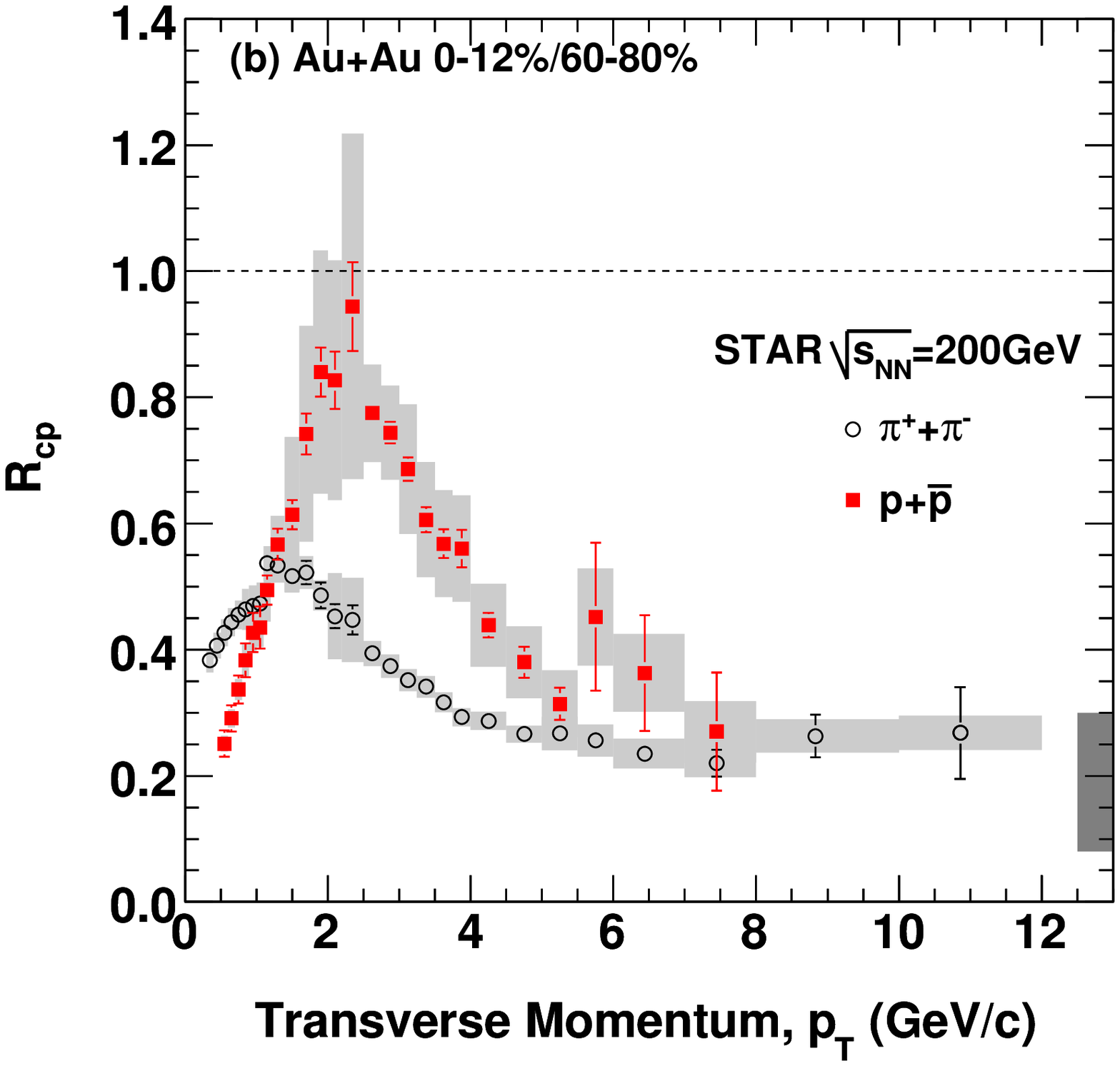 }}
\end{minipage}
\caption{Panel (a) shows the transverse momentum dependence of the
baryon-to-meson ($\overline{p}/\pi^{-}$) ratio in central (0-12\%,
circles) and peripheral (60-80\%, squares) Au+Au collisions
at $\sqrt{s_{NN}})=200$~GeV .  Data from p+p (shaded bands) and
d+Au (triangles) collisions are shown for reference.
Panel (b) shows the spectral modification (R$_{CP}$) for pions (open
circles) and protons (closed squares) for central (0-12\%) relative to
peripheral (60-80\%) collisions.  The light shaded bands represent the 
point to point systematic uncertainty.  The darker shaded band represents
the normalization systematic uncertainty in the number of binary collisions.
\label{fig:BaryonEnchancement}}
\end{figure}

One of the intriguing observations at RHIC is the increase in
the number of baryons relative to mesons in the intermediate-\pt~region
as compared to more elementary collisions, p+p and $e^{+}+e^{-}$.
This enhancement, illustrated in Fig.~\ref{fig:BaryonEnchancement}a, is
seen to depend strongly on the centrality of the collision.  With the
multiple particle identification techniques, the STAR results
completely cover a wide range of transverse-momentum.  The relative
enhancement, which is found to be maximal at \pt$\sim2-3$~GeV/$c$,
is more predominant for the most central\cite{cite:STAR_Id200AuAu}
collisions.  For peripheral\cite{cite:STAR_Id200AuAu} collisions, the
enhancement over p+p\cite{cite:STAR_Id200dAupp} collisions is found to
diminish leaving the data essentially unmodified.  d+Au
data\cite{cite:STAR_Id200dAupp}, at $\sqrt{s_{NN}}=200$~GeV, is found
to exhibit no enhancement relative to the p+p collisions.  For particles
with a higher transverse momenta, the enhancement disappears with the
ratio approaching the p+p reference at \pt$\sim5$GeV$/c$.

If indeed protons are predominantly produced in gluon jets and pions in
quark jets, two possible explanations of these effects can be discussed.
Firstly, one could consider that a gluon jet could be more easily
propagated through the medium than a quark jet, leading to an increase
in the number of protons in the intermediate-\pt~region.  This, however,
contradicts theoretical predictions\cite{cite:Theory_Vitev} where an
opposite effect was expected.  Alternatively, more gluon jets could be
initially produced, or {\it induced}, for the more central data.
The latter appears to be the more plausible, as the highest-\pt~data
exhibits little or no enhancement over the p+p data, indicating a
similar energy loss for gluons and quarks.  This is further
substantiated by the comparison of pion and proton $R_{cp}$
(Fig.~\ref{fig:BaryonEnchancement}b), where no difference is found
in the suppression at high-\pt\cite{cite:STAR_Id200AuAu}.
Alternative approaches to explain the phenomenon observed in the data,
have also been developed.  For example, the recombination/fragmentation
picture of thermal/shower partons has had success at describing this
data\cite{cite:HwaRecombination}.  To distinguish between the
different proposed mechanisms, further differential analysis is needed,
perhaps by selecting on the medium path length via the collision's
reaction plane-dependent analysis.

Additional information on the observed enhancement of baryons has come
from two further sources of systematic study.  Firstly, the RHIC
facility has produced collisions at a reduced center-of-mass energy
of $\sqrt{s_{NN}}=62.4$~GeV.  The measured baryon enhancement is also
observed for this incident energy, although the effect is magnified
for the proton over pion, presumably due to higher baryon transport for
this lower energy.  For the antiproton over pion ratio, the enhancement
is lower due to a smaller number of primordial anti-particles being
produced\cite{cite:STAR_62Id}.  A second systematic study has found that, through colliding
copper nuclei at the same center-of-mass energy, there is no
collision-species dependence of the modification, as long as data with
the same number of participants are compared.  For both of these, the
same systematic \pt~dependences are seen: enhancement over p+p
collisions at intermediate-\pt, and no enhancement at high-\pt.

\section{Conclusions}

Measurements of identified protons and pions from low- to high-\pt~have
proven to be a valuable tool in understanding the particle production
and energy loss mechanisms in relativistic heavy-ion collisions. The
suppression of pions at high-\pt~lead us to conclude that the partons
undergo a large energy loss due to a hot, dense medium created during the
collisions.  Further studies, through the analysis of protons and pions, 
indicate that the partonic energy loss is similar for both the gluons and
quarks.  The amount of energy loss suffered by the partons is found to
be strongly $N_{part}$ dependent.  For different collision species, the
suppression is found to be invariant for the same number of participants.

\section*{Acknowledgements}

We thank the RHIC Operations Group and RCF at BNL, and the
NERSC Center at LBNL for their support. This work was supported
in part by the Offices of NP and HEP within the U.S. DOE Office 
of Science; the U.S. NSF; the BMBF of Germany; CNRS/IN2P3, RA, RPL, and
EMN of France; EPSRC of the United Kingdom; FAPESP of Brazil;
the Russian Ministry of Science and Technology; the Ministry of
Education and the NNSFC of China; IRP and GA of the Czech Republic,
FOM of the Netherlands, DAE, DST, and CSIR of the Government
of India; Swiss NSF; the Polish State Committee for Scientific 
Research; SRDA of Slovakia, and the Korea Sci. \& Eng. Foundation.

\newcommand{\etal} {$\mathrm{\it et\ al.}$}


\begin{thebibliography}{0}
\bibitem{cite:STAR_Id200dAupp} J.Adams {\it et al} Phys. Lett. {\bf B637} (2006) 161
\bibitem{cite:STAR_Id200AuAu} B.I.Abelev {\it et al} Phys. Rev. Lett. {\bf 97} (2006) 152301
\bibitem{cite:STAR_SpectralSuppression} J.Adams {\it et al} Phys. Rev. Lett. {\bf 91} (2003) 172302
\bibitem{cite:STAR_BackToBack} C.Adler {\it et al} Phys. Rev. Lett. {\bf 90} (2003) 082302
\bibitem{cite:STAR_ToF} J.Adams {\it et al} Phys. Lett. {\bf B616} (2005) 8
\bibitem{cite:Theory_Vitev} I.Vitev Phys. Lett. {\bf B639} (2006) 38
\bibitem{cite:STAR_62Id} B.I.Abelev {\it et al} arXiv:nucl-ex/0703040 {\it submitted to Phys. Lett. B}
\bibitem{cite:HwaRecombination} R.C.Hwa and C.B.Yang Phys.Rev.C {\bf 70} (2004) 024905
\end{thebibliography}
\end{document}